\documentclass[10pt,letterpaper]{article}
\usepackage{authblk}

\usepackage[a4paper,top=3cm,bottom=2cm,left=3cm,right=3cm,marginparwidth=1.75cm]{geometry}

\usepackage{threeparttable}

\usepackage[super,comma]{natbib}
\usepackage{siunitx}

\usepackage{cuted}
\usepackage{graphicx}
\usepackage{makeidx}  
\usepackage{multicol}
\usepackage{amsmath,graphicx}
\usepackage{amssymb}
\usepackage{tabulary}
\usepackage{tabularx}
\usepackage{float}
\usepackage{todonotes}
\usepackage{color}
\usepackage[export]{adjustbox}
\usepackage{amsmath}
\usepackage{amssymb}
\usepackage{multirow} 
\usepackage{floatflt}
\usepackage{url}
\usepackage{epstopdf}
\usepackage{xcolor}
\usepackage{algorithm}

\definecolor{darkgreen}{RGB}{0, 190, 0}

\newcommand{\bfB}{\mathbf{B}}
\newcommand{\bfr}{\mathbf{r}}

\newcommand{\bfv}{\mathbf{v}}

\newcommand{\bfM}{\mathbf{M}}
\newcommand{\mdiag}{\mathrm{{diag}}}

\newcommand{\bfa}{\mathbf{a}}

\newcommand{\bfrho}{\boldsymbol{\rho}}
\newcommand{\bfRho}{\mathbf{P}}
\newcommand{\Nc}{N_\text{c}}
\newcommand{\Nv}{N_\text{v}}
\newcommand{\Ns}{N_\text{s}}

\newcommand{\Nt}{N_\text{t}}

\newcommand{\bbCo}{\mathbb{C}}
\newcommand{\defeq}{\overset{\underset{\mathrm{def}}{}}{=}}

\newcommand{\bfPhi}{\boldsymbol{\Phi}}

\newcommand{\argmin}{\operatornamewithlimits{argmin}} 
\newcommand{\matr}[1]{\mathbf{#1}}
\newcommand{\vectr}[1]{\mathbf{#1}}
\newcommand*{\tran}{^{\mkern-1.5mu\mathsf{T}}}
\newcommand*{\conj}{^{\mkern-1.5mu\mathsf{H}}}
\newcommand{\vvv}[1]{\textcolor{black}{{#1}}}
\newcommand{\vvvv}[1]{\textcolor{black}{{#1}}}

\newcommand{\headrow}{}
\newcommand*{\thead}[1]{\multicolumn{1}{c}{\bfseries #1}}

\title{\textrm{Deep variational network for rapid 4D flow MRI reconstruction}}

\author[1*]{Valery~Vishnevskiy}
\author[1*]{Jonas~Walheim}
\author{Sebastian~Kozerke}

\affil{Institute for Biomedical Engineering, University and ETH Zurich, Zurich, Switzerland}
\affil[*]{Equal contribution}

\begin{document}

	\maketitle
	
	\begin{abstract}
		Phase-contrast (PC) magnetic resonance imaging provides time-resolved quantification of blood flow dynamics that can aid clinical diagnosis.
		Long in vivo scan times due to repeated 3D volume sampling over cardiac phases and breathing cycles necessitate accelerated imaging techniques that leverage data correlations. 
		Standard compressed sensing reconstruction methods require tuning of hyperparameters and are computationally expensive, which diminishes potential reduction of examination times.
		We propose an efficient model-based deep neural reconstruction network and evaluate its performance on clinical aortic flow data.
		\vvvv{The network is shown to reconstruct undersampled 4D flow MRI data in under a minute on the standard consumer hardware.}
		Remarkably, the relatively low amounts of tunable parameters allowed the network to be trained on images from 11 reference scans while generalizing well to retrospective and prospective undersampled data for various acceleration factors and anatomies. 
	\end{abstract}

\subsection*{Introduction}
4D flow MRI provides spatio-temporally resolved quantification of blood flow and offers great potential for the assessment of cardiovascular disease, e.g. aortic valve stenosis, atherosclerosis, or vessel wall remodelling \cite{markl20124d}. 
However, clinical adaptation of the method has been hampered by long exam times. 

Many efforts have been dedicated to accelerate flow acquisition by exploiting redundancies in the data. 
Partial Fourier imaging \cite{feinberg1986halving} has been used for moderate acceleration \cite{szarf2006zero}, but the underlying assumption of a slowly varying phase has been shown to be incorrect for 4D flow MRI \cite{walheim2019limitations}. 
Parallel imaging (PI) \cite{pruessmann1999sense}, which exploits the spatially varying sensitivity of receiver elements in the coil array has become a standard for accelerated imaging, but undersampling rates are limited by noise amplification~\cite{wiesinger2004electrodynamics}. 
The advent of compressed sensing (CS)~\cite{lustig2007sparse} has enabled  acceleration of 4D flow MRI by acquiring only a subset of k-space data and exploiting prior information about data regularities during reconstruction \cite{kim2012accelerated,valvano2017accelerating,bollache2018k,walheim20195dflow,ma2019aortic,rich2019bayesian}, with typical acceleration factors ranging from 5 \cite{bollache2018k} up to 27 \cite{rich2019bayesian}. 
In particular, the locally low rank (LLR) regularized reconstruction~\cite{zhang2015accelerating} has been a successful technique, which iteratively balances the data fidelity cost and the singular norm of a patch matrix stacked over cardiac phases (see Methods for details).
However, iterative reconstruction methods as used in CS increase reconstruction times considerably, implying that evaluation of 4D flow MRI data will typically happen when the subject has already been moved out of the scanner. 

In recent years, deep neural networks have gained increasing popularity in MR image reconstruction. 
In the training stage, the neural network learns abstract features from a set of scans. 
After training, newly acquired data are reconstructed with very little computational effort by inference with the learned weights. 
This reduction in reconstruction times can facilitate the use of accelerated imaging methods in clinical practice. 
Moreover, reconstruction results can be superior to traditional CS methods \cite{hammernik2018learning,mardani2018deep}.
Some approaches discard concepts of iterative image reconstruction altogether, e.g. by learning end-to-end mappings from k-space to image space~\cite{zhu2018image}. 
As a downside, such networks usually require abundant amounts of high quality training data which is not available for high-dimensional flow MRI.
Model-based neural reconstruction networks can also be designed to replicate the behaviour of an iterative reconstruction by interlacing nonlinear convolutional filters with an operation that enforces closeness of the current image estimate to the acquired data~\cite{hammernik2018learning, schlemper2017deep,mardani2018deep, maier2019learning}, similar to the data fidelity step in an iterative shrinkage-thresholding algorithm~\cite{beck2009fast}.
A recent study \cite{antun2019instabilities} showed that neural network architectures which incorporate such an operation generalize better to different undersampling rates. 
In contrast, generic architectures which are solely based on convolutional layers can even lead to deteriorated image quality when the undersampling factor is increased, although one would expect the reconstruction result to improve when more information is available. 
\vvv{The adversarial approach for training MR reconstruction networks~\cite{yang2017dagan, quan2018compressed} is usually aimed at improving perceptual reconstruction quality, such as image sharpness. 
	Typically, this is achieved at the expense of reconstruction nRMSE (normalized root-mean-square error)~\cite{narnhofer2019inverse}, which is critical for flow quantification.}

In this work, an approach based on the idea of deep variational neural networks~\cite{hammernik2018learning} is implemented for rapid 4D flow reconstruction, wich is referred to as FlowVN hereafter.. 
\vvv{For comparsion, the 3D variational network (VN) architecture as presented in~\cite{hammernik2018learning} was adapted for 4D flow data by using 3D filter banks operating on subsets of $xyzt$ dimensional data, yielding the model that we refer to as HamVN.}
The FlowVN twork architecture replicates 10 steps of an iterative image reconstruction, while allowing for learnable spatio-temporal filter kernels, activation functions, and regularization weights in each iteration. 
It is demonstrated that based on training performed with retrospectively undersampled data of healthy subjects, FlowVN can accurately reconstruct pathological flow in a stenotic aorta in 21 seconds. 
Moreover, an imaging study with healthy subjects demonstrates good agreement of reconstructions from prospective undersampling with reference measurement.
\begin{figure*}[th!]
	\centering
	\includegraphics[width=0.9\textwidth]{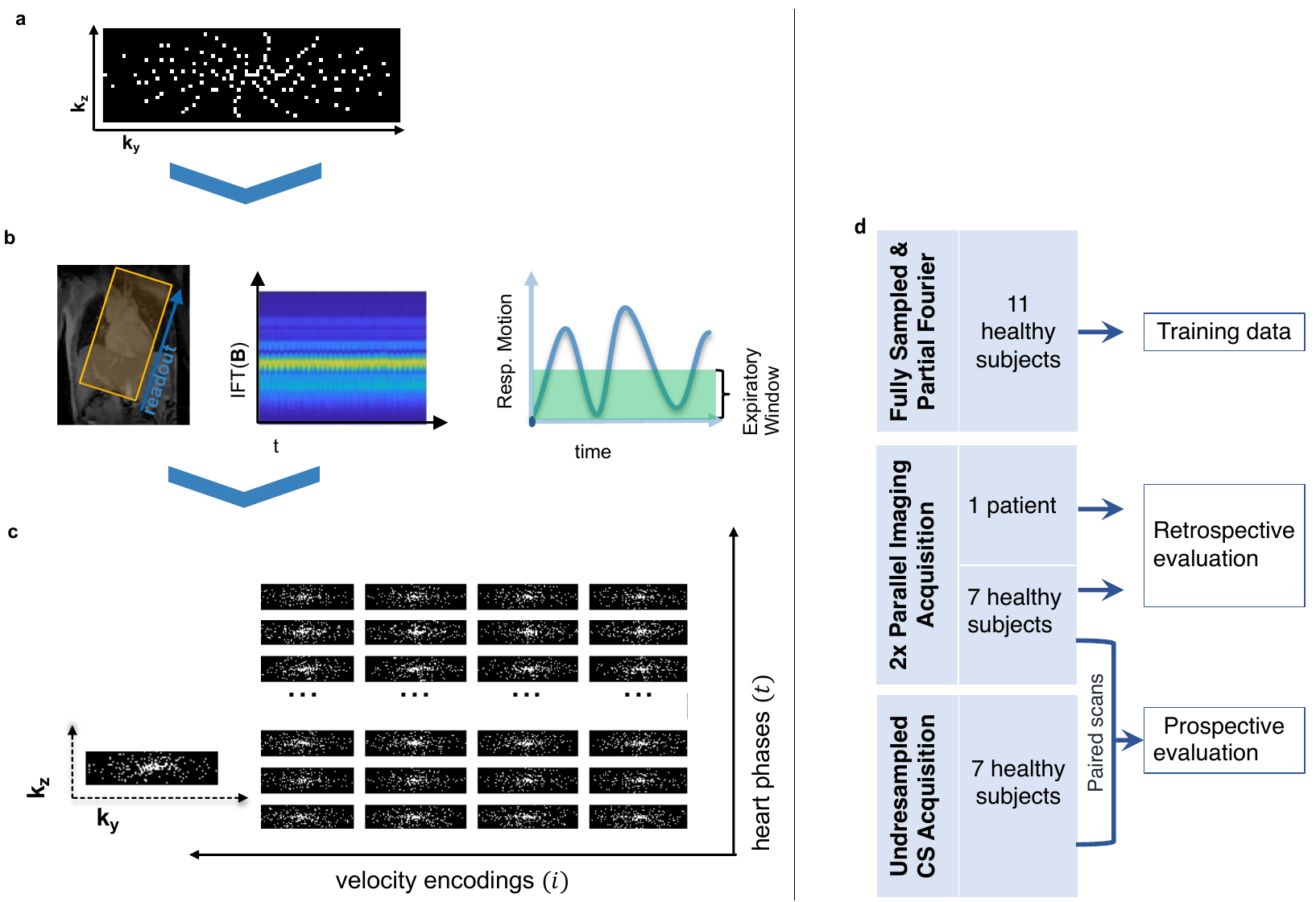}
	\caption{\textbf{Breathing resolved 4D flow data acquisition. }
		\textbf{a}, Data are sampled using a Cartesian pseudo-radial tiny Golden angle sampling pattern. 
		\textbf{b}, Time-resolved data are binned to end expiration using a combination of principal component analysis, low-pass filtering, and coil-clustering. 
		\textbf{c}, Acquired data are sorted according to heart phase and velocity encoding.
		\textbf{d}, Datasets used during training and evaluation.
		\vvv{To conduct prospective evaluation, 7 healthy volunteers underwent accelerated CS and reference PI acquisitions during the same scan session.}
	}
	\label{fig:cs_acq_data}
\end{figure*}

\subsection*{4D flow MRI reconstruction with FlowVN}
The FlowVN architecture improves HamVN in the following ways: 
\vvv{(i) linear activations are used instead of radial basis functions (RBF)}, (ii) the network is conditioned on the sampling rate, (iii) exponential weighting of intermediate layers is used as regularization, (iv) real and imaginary parts of the signal are filtered by shared weights, (v) momentum is considered during gradient descent (GD) unrolling, and, (vi) the data term allows tunable activation functions.
The network is trained for a wide range of acceleration factors by allowing acceleration dependent weighting of data consistency and filtering steps.

As illustrated in Fig.~\ref{fig:cs_acq_data}a, for each velocity encoding direction, the k-space data is acquired using a Cartesian Golden angle sampling strategy, yielding variable density undersampling patterns in k-space.
The signal of a total of 28 physical coils is compressed into 5 virtual coils via clustering~\cite{zhang2010magnetic}.
The samples are then sorted into respiratory bins and data in the end-expiratory bin is used for reconstruction.

\begin{figure*}[t!]
	\centering
	\includegraphics[width=0.99\linewidth]{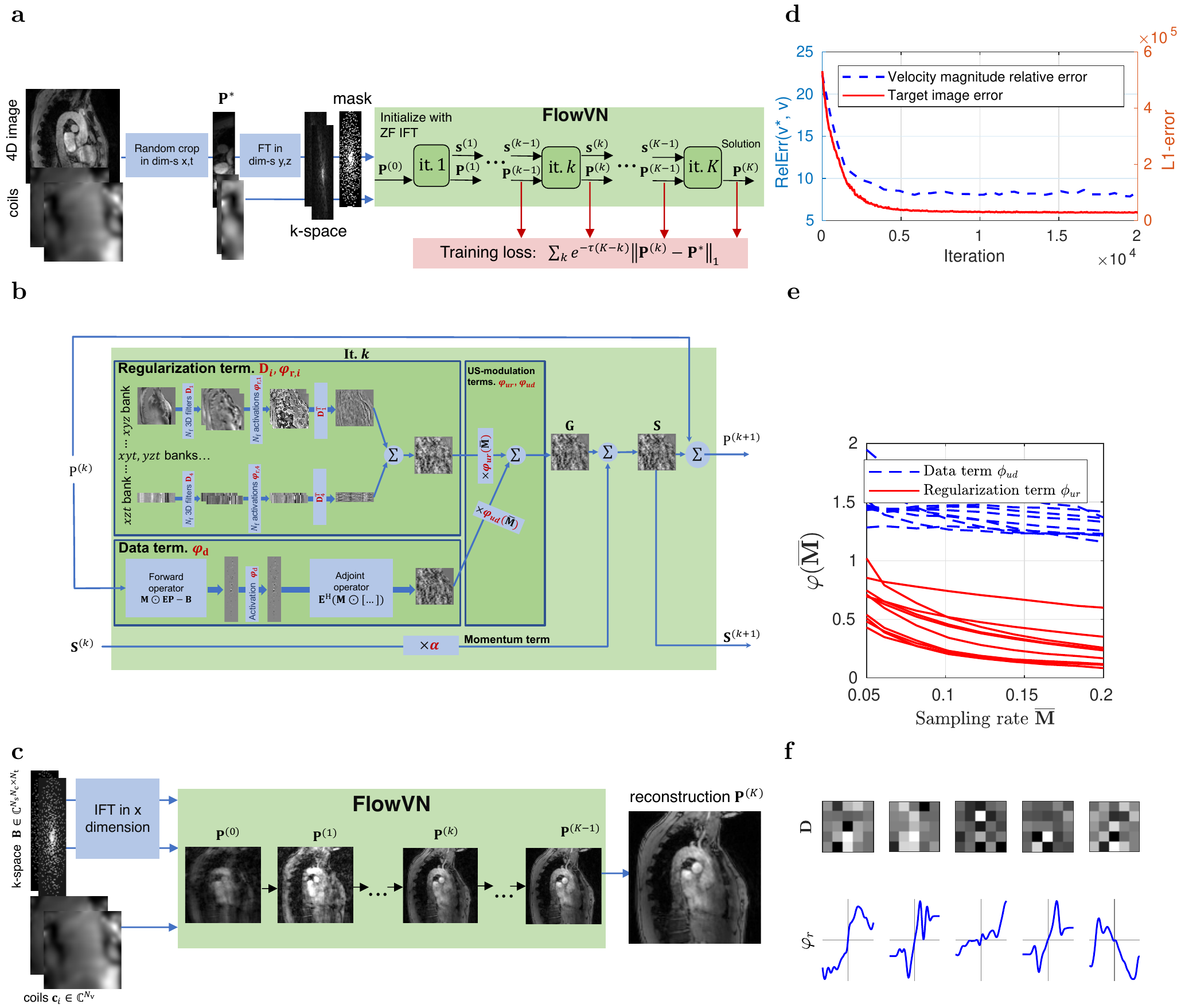}
	\caption{
		\textbf{FlowVN architecture and training.}
		\textbf{a}, Structure of FlowVN and its training strategy which uses a reduced  field-of-view data.
		\textbf{b}, Single unrolled iteration block \vvv{consisting of convolutional filtering, data fidelity, undersampling modulation and gradient descent momentum terms. 
			Tunable parameters are highlighted in red.}
		\textbf{c}, FlowVN at inference time yields 4D image reconstruction.
		\textbf{d}, Target training image error and velocity magnitude error in the aorta \vvv{evaluated using training data}.
		\textbf{e}, Data and gradient term weighting functions shown for each of 10 layers.
		\textbf{f}, Exemplary slices of 3D $xyz$ filters and their corresponding activation functions at layer 5.
	}
	\label{fig:sketch_train}
\end{figure*}

A deep variational network can be seen as a differentiable sequence of an unrolled numerical optimization scheme.
To enable learning, such sequence is then relaxed by allowing tunable filter weights and activation functions. 
As described in Methods, we unroll $K$=$10$ steps of a gradient descent with momentum governed by a scalar $\alpha^{(k)}$:
\begin{align}
&\vectr{S}^{(k+1)}\leftarrow \alpha^{(k+1)}\vectr{S}^{(k)} + \vectr{G}^{(k)},\\
&\bfRho^{(k+1)}\leftarrow \bfRho^{(k)} - \vectr{S}^{(k+1)}.
\end{align}
At each $k$-th layer $\bfRho^{(k)}$ the current complex-valued spatio-temporal image estimate is represented, while $\vectr{S}^{(k)}$ maintains a running average of update steps.
The update step $\vectr{G}^{(k)}$ consists of the data consistency and regularization terms (see Methods and Supplementary Algorithm~1 for details), that are weighted according to the sampling rate \vvv{$\overline{M}=1/R$ ($R$ is the acceleration factor)} via tunable activation functions \vvv{$\varphi_\mathrm{ud}^{(k)}(\overline{M})$ and $\varphi_\mathrm{ur}^{(k)}(\overline{M})$, respectively}.
The data consistency term modulates the k-space data residual via an activation function and maps them back to the image space via a conjugate imaging operator.
The regularization term at each layer contains 3D filters grouped into 4 banks, where each bank performs convolutions in 3 dedicated dimensions, namely $xyz$, $xyt$, $xzt$ and $yzt$, therefore avoiding costly 4D convolutions.
To avoid overfitting, we assume shared filters and activation functions that operate on real and imaginary components of the image.
Note that both data and regularization terms do not assume correlations between real and imaginary parts of the signal, as highlighted in Fig.~\ref{fig:sketch_train}b.

The image estimate $\bfRho^{(K)}(\bfB, \Theta)$ of the final layer can be then seen as a function of the k-space samples $\bfB$ and network parameters $\Theta$.
To tune the network parameters $\Theta$ we minimize the layer-wise exponentially weighted $\ell_1$ image reconstruction loss:
\begin{equation}
\label{eq:train_exp_loss}
\min_\Theta \mathop{\mathbb{E}}_{ \{\bfB, \bfRho^\ast\} \sim \mathcal{T}}
\sum_{k=1}^K
e^{-\tau(K-k)}\,\| \bfRho^{(k)}(\bfB; \,\Theta) - \bfRho^\ast\|_1,
\end{equation}
over the retrospectively undersampled training dataset $\mathcal{T}$, where
$\bfRho^\ast$ is the ground truth image. 
Layer weighting is controlled by parameter $\tau$$\geq$$0$: when $\tau\approx0$, the reconstruction error is penalized equally across layers, therefore gradients of network parameters have lower variance during stochastic optimization, yielding faster convergence.
On the contrary, when $\tau$$\rightarrow+$$\infty$, only reconstruction at the final layer $\bfRho^{(K)}$ is minimized, which improves fitting accuracy on the  training data.
It is worth mentioning that $\tau$ controls the trade-off between training reconstruction residual and network regularity.
Similarly to Landweber iterations~\cite{landweber1951iteration, hammernik2018learning} and deep supervision~\cite{liu2016learning}, such implicit regularization penalizes irregular representations at intermediate layers and favors networks that can provide fast reconstruction. 
We propose to initialize $\tau$ with zero and then gradually increase it according to the training schedule (see Methods).

\begin{table}[bt]
	\caption{
		\textbf{Model complexities and typical reconstruction time for 4D flow reconstruction.}
		Typical reconstruction time for 4-point velocity encoded data compressed to 5 virtual coils and reconstructed on 113$\times$113$\times$25 grid. 
		CS-LLR was executed on a 6-core Intel CPU, FlowVN \vvv{and HamVN were}  implemented in Tensorflow and evaluated on a NVIDIA Titan RTX.
	}
	\label{tbl:run_times}
	\centering
	\begin{threeparttable}
		\begin{tabular}{llc}
			\headrow
			\hline
			\thead{Method} & \thead{Recon. Time} & \thead{\# of Param.} \\
			\hline
			\shortstack[l]{
				{CS-LLR}
			} 
			& 
			\shortstack[l]{
				{\color{red}\rule{100pt}{8pt}}\\
				10\,min\,24\,s
			}
			&
			2
			\\ \hline
			\shortstack[l]{
				\vvv{HamVN}
			}  &
			\shortstack[l]{
				{\color{red}\rule{6.25pt}{8pt}}\\
				\vvv{89\,s}
			} 
			&
			\vvv{62,742}
			\\ \hline
			\shortstack[l]{
				{FlowVN}
			}  &
			\shortstack[l]{
				{\color{red}\rule{3.36pt}{8pt}}\\
				21\,s
			} 
			&
			63,583
			\\
			\hline
		\end{tabular}
	\end{threeparttable}
\end{table}

To demonstrate the validity of our approach, we note that the extracted velocity magnitude error in the aorta decreases simultaneously with the target reconstruction error \vvvv{during training}, as shown in Fig.~\ref{fig:sketch_train}d, therefore indicating that \vvv{the target $\ell_1$ image error is a valid training surrogate.}
It can be seen from Fig.~\ref{fig:sketch_train}e that the regularization term is suppressed for lower acceleration factors $R$ (higher sampling rate $\overline{\bfM}$).
A subset of learned FlowVN parameters $\Theta$ is shown in Fig.~\ref{fig:sketch_train}f illustrating that learned convolutions perform direction-dependent filtering.

\subsection*{Retrospective and Prospective Evaluation}
\begin{figure*}[t]
	\includegraphics[width=\linewidth]{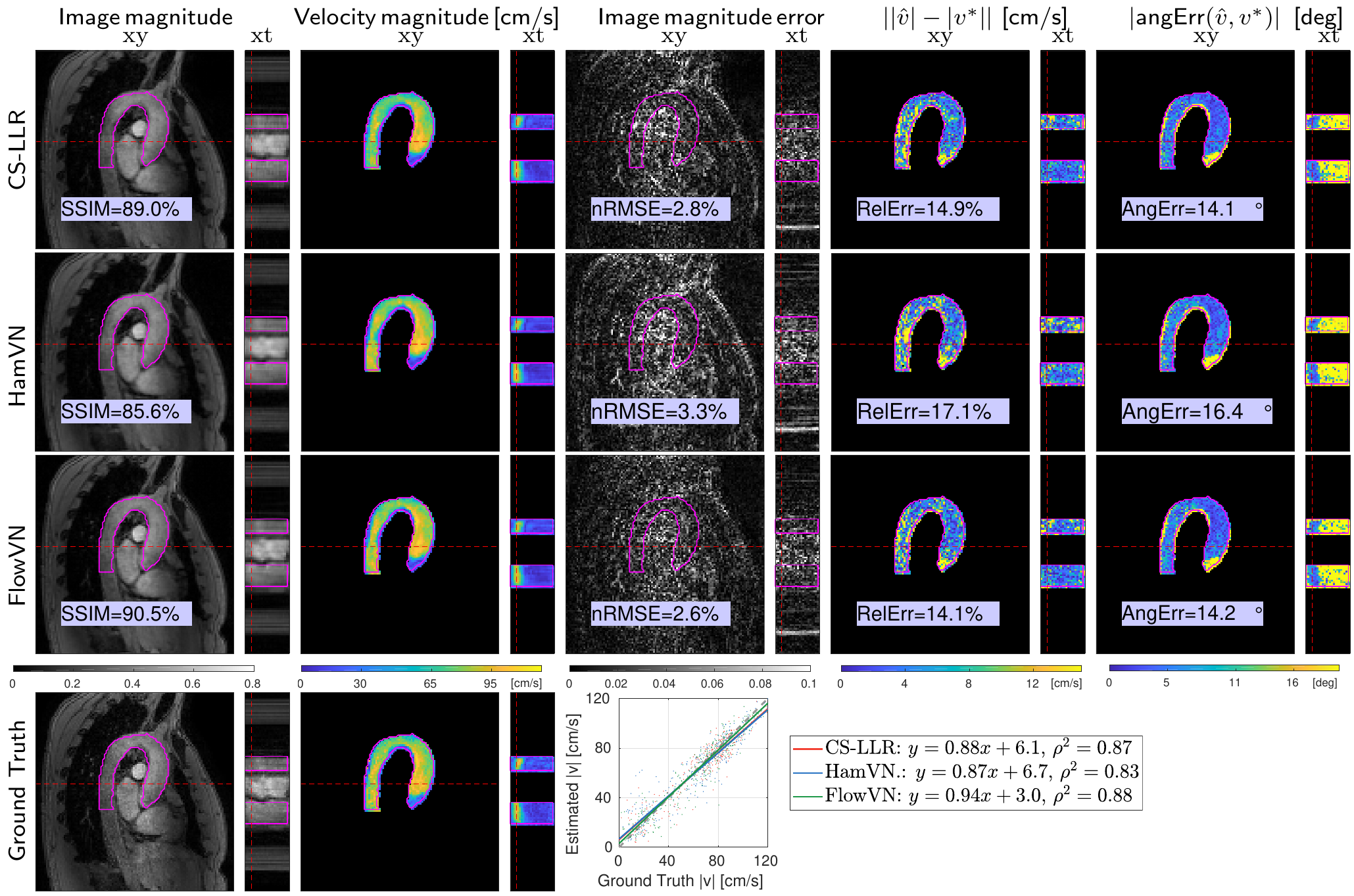}
	\caption{
		\textbf{Reconstruction results on retrospectively undersampled data.}
		Image magnitudes and estimated 4D velocity magnitude maps on retrospectively 14$\times$ undersampled data from a healthy volunteer.
		Corresponding slice locations are illustrated with red dashed lines, indicating crossection of the aorta and systolic peak.
		Scatter plot of velocity magnitude over manually segmented aorta (contour shown in magenta) is given together with correlation analysis ($y=ax+b$).
	} 
	\label{fig:bigimg}
\end{figure*}

Reconstructed image magnitudes (for a single velocity encoding component), estimated velocity magnitudes and their errors of a healthy volunteer data for acceleration factor $R$=14 \vvvv{are illustrated in Fig.~\ref{fig:bigimg} for retrospectively undersampled data}.
Compared to CS-LLR and HamVN, the proposed FlowVN provides better reconstruction accuracy in terms of image magnitude and velocities.
Scatter plot and correlation analysis further suggest that the velocity magnitude image estimated via FlowVN is in better agreement with ground truth. 
As shown in Supplementary Table~\ref{tbl:retro_errors} these observations extend to other acceleration factors $R$ (6--22) as tested on 7 healthy volunteers.

\begin{figure*}[t]
	\includegraphics[width=0.8\linewidth]{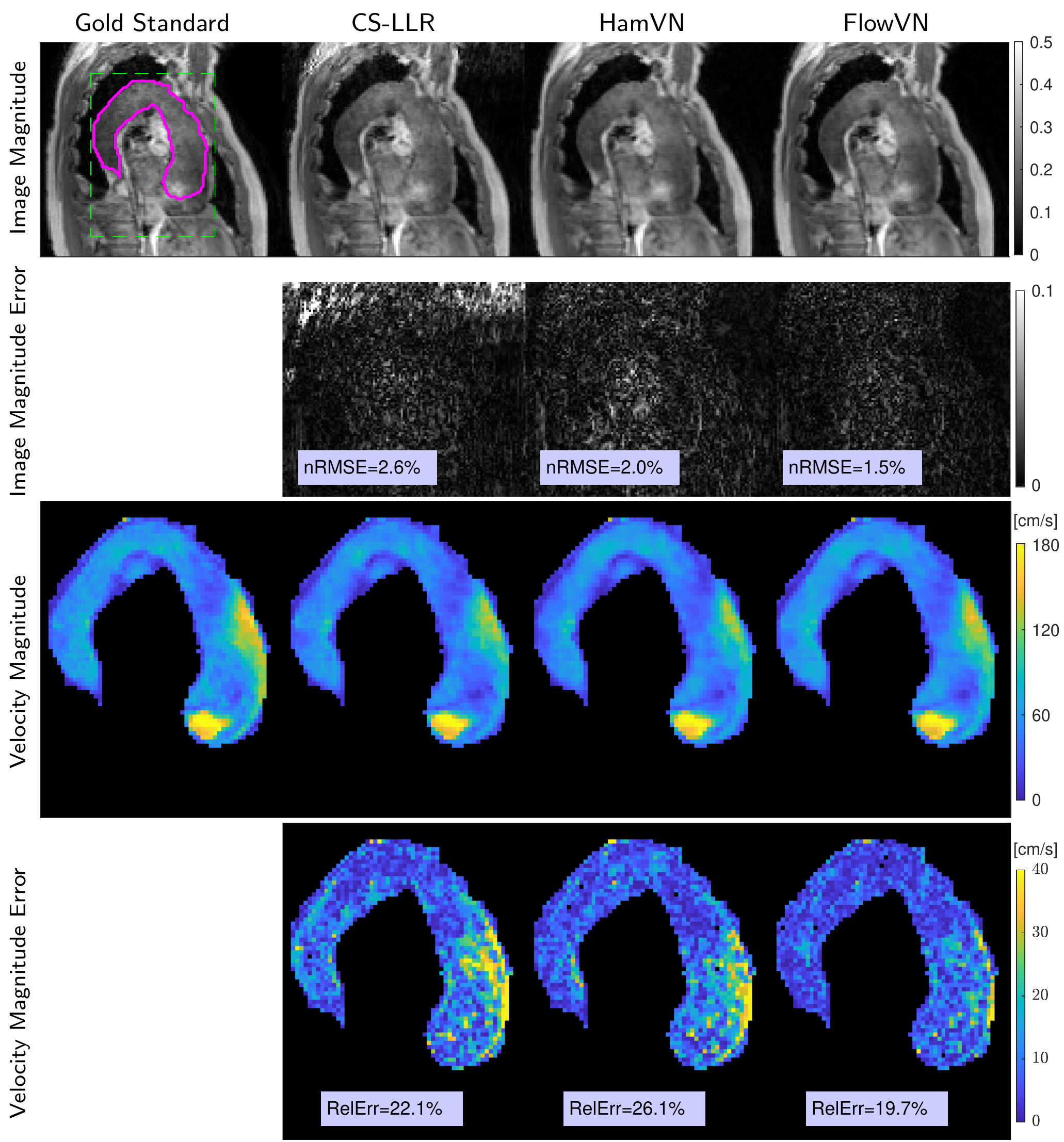}
	\caption{
		\textbf{Retrospective reconstruction of the data from patient with abnormal flow pattern.}
		Reconstruction results of 10$\times$ retrospectively undersampled patient data shown at systolic peak flow.
		Manual aorta segmentation and field-of-view are shown with magenta and green dashed lines, respectively.
	} 
	\label{fig:patient}
\end{figure*}

Fig.~\ref{fig:patient} indicates that FlowVN can accurately reconstruct the jet at the inlet section of the aorta for a patient with a pathological aortic valve.

\begin{figure*}[bt]
	\includegraphics[width=\linewidth]{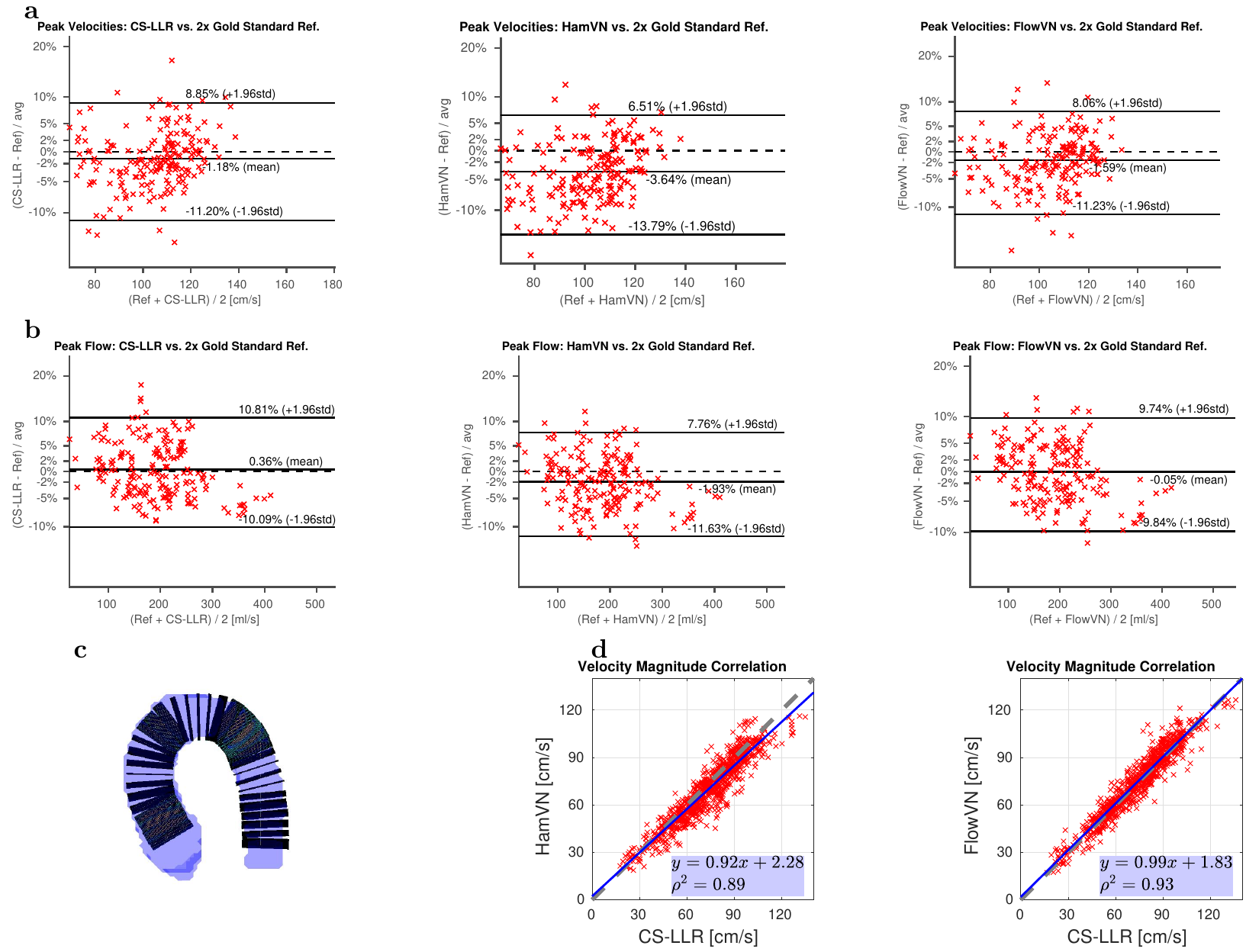}
	\caption{
		\textbf{Quantitative flow evaluation of reconstruction methods on the prospectively undersampled data ($12.4\leq R\leq13.8$) \vvvv{from 7 healthy volunteers}.}
		\textbf{a,b}, Bland-Altman analysis of peak velocities and peak flow assessed over manually segmented aorta slices as illustrated in \textbf{c}. 
		\vvvv{Black dashed bars indicate zero difference between corresponding measurements and solid lines show mean and standard deviation values.}
		\textbf{d}, Correlation analysis of aortic velocity magnitude estimates by learned VN architectures and CS-LLR reconstruction. 
		\vvvv{The blue line shows the linear regression fit and grey dashed line corresponds to $y=x$.}
	}
	\label{fig:pro_flow}
\end{figure*}

\vvvv{The prospective undersampling acquisition results are} reported in Fig.~\ref{fig:pro_flow}a,b: peak velocities and peak flow estimated using CS-LLR and FlowVN are in good agreement with PI reconstruction, while HamVN systematically underestimates velocity magnitudes.
Moreover, correlation analysis in Fig.~\ref{fig:pro_flow}d reveals high correlation between CS-LLR and FlowVN velocity estimates.
In contrast, HamVN shows systematic velocity underestimation, compared to CS-LLR.

The exemplary reconstruction time for typical 4-point velocity encoded images reported in Table~\ref{tbl:run_times} shows that the proposed FlowVN is 30 times faster than CS-LLR reconstruction.

\subsection*{Discussion}
Practical learning-based image reconstruction can be traced back to dictionary learning methods~\cite{ravishankar2010mr, caballero2014dictionary}, where prior information is learned from image patches and then used as a sparsity-inducing regularizer for iterative reconstruction.
Such an approach yields orders of magnitude longer reconstruction times, compared to modern deep learning approaches.
\vvv{A straightforward application of deep artificial neural networks has been suggested for learning} reconstruction as a regression from the k-space~\cite{zhu2018image} or zero-filled reconstructions~\cite{lee2017deep} directly into the image space.
Although tempting, such an approach might be unjustified, because the k-space and zero-filling artefacts have global dependence on image intensities. 
The advent of effective automatic differentiation systems~\cite{lecun1988theoretical, abadi2016tensorflow} revitalized the idea of unrolling~\cite{domke2012generic} and relaxing numerical schemes that can solve the original reconstruction problem.
Following this approach, a number of deep neural network architectures were proposed~\cite{schlemper2017deep, sun2016deep, hammernik2018learning, jin2017deep}, that disentangle image acquisition and image prior models.
Unrolling gradient descent reconstruction with tunable filters and activation functions yields the HamVN architecture proposed by Hammernik et al.~\cite{hammernik2018learning}.
One advantage of VNs is that, compared to other deep architectures, they employ relatively limited number of free parameters to tune, therefore they are less susceptible to overfitting.

\begin{figure*}[h!]
	\centering
	\includegraphics[width=0.8\linewidth]{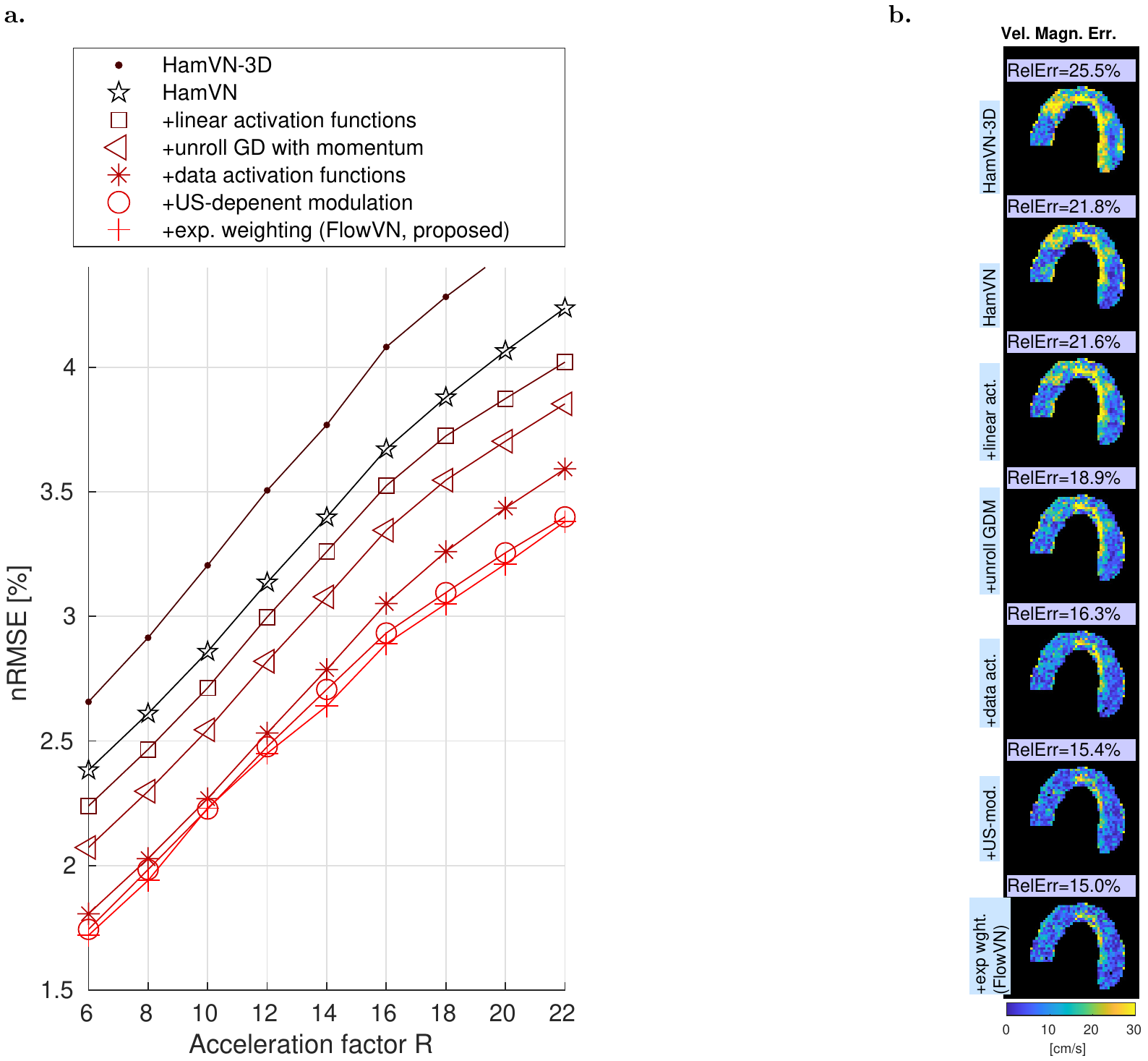}
	\caption{
		\textbf{Retrospective ablation study.} 
		\textbf{a},
		Comparison of VN architectures and modifications proposed in this paper, using retrospectively undersampled data from 7 healthy volunteers for different acceleration factors $R$.
		HamVN-3D performs 3D filtering of $yzt$ dimensional data (assuming that data is fully sampled in the $x$ dimension).
		Modifications are accumulated as presented in the legend and, eventually, yield the FlowVN configuration.
		Undersampling-dependent modulation improves reconstruction quality only for higher acceleration factors.
		Note that the layer-wise exponential weighting of the reconstruction loss does not increase model complexity and, therefore, does not improve reconstruction accuracy for retrospective acceleration simulations.
		\textbf{b},
		\vvvv{Velocity magnitude error maps for $R$=16 shown at systolic peak flow.}
	}
	\label{fig:vol_ablation}
\end{figure*}

In this work we have further developed the VN architecture~\cite{hammernik2018learning, vishnevskiy2018image, vishnevskiy2019deep} to accommodate high performance undersampled 4D flow reconstruction with limited training. 
Namely, we avoid exponential model complexity growth by avoiding 4D convolutions and by using separable 3D convolutions that are shared for real and imaginary parts of the image.
Furthermore, in contrast to the original HamVN~\cite{hammernik2018learning}, we train our FlowVN for a wide range of undersampling factors by allowing the regularization term to depend on them.
As illustrated in Fig.~\ref{fig:sketch_train}e, regularization scaled by $\varphi_\text{ur}$ decreases as more samples are available, while the data term $\varphi_\text{ud}$ stays constant for most of the layers.
Such conditioning allows network training on a larger variety of artefacts and as it is necessary in practice, since for a given fixed acquisition time, the precise value of the undersampling factor is not known a priori and depends on breathing and cardiac motion patterns.
We hypothesize that the wide range of acceleration factors which were used simultaneously to train the FlowVN provided a diverse collection of aliasing artefacts and enabled robust learning on a remarkably limited training set of 11 subjects.
The exponential weighting of the layer-wise reconstruction loss~(\ref{eq:train_exp_loss}) further regularized FlowVN parameters by penalizing the nonlinear behaviour presented in HamVN reconstructions.
\vvv{Supplemental Fig.~\ref{fig:vol_ablation} and Table~\ref{tbl:pro_ablation} provide quantification of reconstruction accuracy effects attributed to the modifications proposed with FlowVN. 
	In particular, modifications to the network architecture result in a model that can better adapt to data and yield higher accuracy for retrospectively undersampled experiments, while the proposed exponential weighting of the training loss improves accuracy of the prospective evaluation, which indicates better generalization ability.
	It is worth noting, that FlowVN has only 1\% more tunable parameters compared to HamVN (c.f. Table~\ref{tbl:run_times}), while improving reconstruction nRMSE by 23\% (averaged over acceleration factors as given in Supplemental Fig.~\ref{fig:vol_ablation}). 
	We note that 4D flow MRI greatly benefits from using coil information during reconstruction (see Supplementary Table~\ref{tbl:pro_ablation}).
	Accordingly, comparison with single-coil reconstruction networks~\cite{schlemper2017deep} has limited benefit.
}

The proposed FlowVN is a learning-based approach for reconstructing undersampled 4D flow MRI data in under a minute. 
For fixed reconstruction accuracy, FlowVN enables higher acceleration factors (12\% improvement compared to CS-LLR image nRMSE at $R$=16) and does not introduce significant bias of peak flow estimates.
The proposed reconstruction is 30 times faster than state-of-the art CS-LLR \vvv{and 4.2 times faster than HamVN, due to using linear activation functions rather than RBFs, which requires computation of pairwise distances between control knots and image intensities.}
It is worth noting that FlowVN demonstrates high generalization ability, being able to preserve patient pathologies that were not present in the training data.

\clearpage
\newpage
\section*{Methods}
\subsubsection*{Compressed Sensing 4D Flow Reconstruction}
PC MRI encodes flow velocity $\bfv(\bfr, t)\in\mathbb{R}^3$ at spatial location $\bfr$ during cardiac phase $t$ ($1\leq t\leq\Nt$) according to the following equation:
\begin{equation}
\rho_i(\bfr, t)= \rho_0(\bfr, t)\exp\left(j\pi \frac{\left(\bfPhi\mathbf{v}(\bfr, t)\right)_i}{v_\text{enc}} \right),
\end{equation}
where ${v_\text{enc}}$ is the velocity corresponding to a phase of $\pm\pi$, $y_i, i=0,\dots,3$ are the encoded velocity vector components. 
The four-point velocity encoding matrix is given as
\begin{equation}
\bfPhi=
\begin{bmatrix}
0 & 0 & 0 \\
1 & 0 & 0 \\
0 & 1 & 0 \\
0 & 0 & 1 
\end{bmatrix}.
\end{equation}
Therefore, flow velocity $\bfv$ can be calculated from the phase difference of reconstructed PC images $\rho_i$.

Let $\bfrho_{it}\in\bbCo^{\Nv}$ be a discretized image on a $N_x$$\times$$N_y$$\times$$N_z$$=$$\Nv$ grid corresponding to a cardiac phase $t$ and velocity encoding $i$.
Assuming Cartesian sampling on a regular $N_1$$\times$$N_2$$\times$$N_3$$=$$\Ns$ grid, the Fourier transform  $\mathbf{F}\in\bbCo^{\Ns\times\Nv}$, 
and $\Nc$ coil sensitivity maps 
$\mathbf{W}_k\defeq\mdiag(\mathbf{c}_k)\in\bbCo^{\Nv\times\Nv}$ define the spatial encoding operator $ \mathbf{E}\in\mathbb{C}^{\Ns\Nc\times\Nv}$:
\begin{equation}
\mathbf{E}\bfrho \defeq \left[ \left(\mathbf{F}\mathbf{W}_1\bfrho\right)\tran,\dots, \left(\mathbf{F}\mathbf{W}_{N_\text{c}}\bfrho\right)\tran\right]\tran
\in\bbCo^{\Ns\Nc}.
\end{equation}
Considering a single velocity-encoded image sequence, let $\bfRho\in\bbCo^{\Nv\times\Nt}$ and $\mathbf{B}\in\bbCo^{\Ns\Nc\times\Nt}$ be stacked column-vectors of signals $\bfrho$ and zero-filled k-space samples respectively, while $\mathbf{M}\in\{0,1\}^{\Ns\Nc\times\Nt}$ defines the undersampling mask.
Iterative image reconstruction methods seek for a maximum a posteriori (MAP) solution defined by the following optimization problem:
\begin{equation}
\label{eq:opt_gen}
\hat{\bfRho}_\text{MAP}=\argmin_{\bfRho} \frac12\|\mathbf{M}\odot\left(\mathbf{E}\bfRho-\mathbf{B}\right)\|_\text{F}^2 + \mathcal{R}(\mathbf{\bfRho}),
\end{equation}
where the regularization term $\mathcal{R}$ enforces prior assumptions about image regularities.
Herein we consider the local low-rank (LLR) regularization~\cite{zhang2015accelerating} to leverage image correlations among cardiac phases:
\begin{equation}
\label{eq:llr_reg}
\mathcal{R}_\mathrm{LLR}(\bfRho)=\lambda_\mathrm{LLR}\sum_{i\leq N_\mathrm{ptch}}\|\mathbf{T}_i\bfRho\|_\ast,
\end{equation}
where $\mathbf{T}_i\in\{0,1\}^{p^3\times\Nv}$ is the corresponding $p$$\times$$p$$\times$$p$ patch extraction operator, yielding $N_\mathrm{ptch}$  patches, and $\|\cdot\|_\ast$ is the nuclear norm.
For LLR regularization the optimization
problem~(\ref{eq:opt_gen}) is convex and can be efficiently solved using operator splitting techniques such as the fast iterative shrinkage-thresholding algorithm (FISTA)~\cite{beck2009fast}.

\subsubsection*{FlowVN Training}
We employ a $K$=10 layer VN and perform 5$\cdot$$10^4$ iterations of the ADAM algorithm (learning rate $10^{-3}$, $\beta_1$$=$$0.85$, $\beta_2$$=$$0.98$, batch size of 3) for training, during which we continually adjust $\tau=i_\text{opt}$$\cdot$$10^{-3}$ with $i_\text{opt}$ being the iteration number.
On every layer, each 3D filter bank contains $N_\mathrm{f}$=$8$ filters of size $n_\mathrm{c}$=$5$ voxels.
Activation functions $\varphi\{.\}$ are parametrized by $N_\mathrm{knts}$=$91$ control knots with spacing $\omega$=$0.17$:
\begin{equation}
\begin{aligned}
\label{eq:act_fun_def}
\varphi\{h\}=&(1 - h/\omega + \lfloor h/\omega\rfloor)\phi_{\lfloor h/\omega\rfloor}+\\
&(h/\omega - \lfloor h/\omega\rfloor)\phi_{\lfloor h/\omega\rfloor+1},
\end{aligned}
\end{equation}
with gradients provided by the following formulas:
\begin{equation}
\begin{aligned}
&\frac{\partial\varphi}{\partial \phi_i}\{h\}=
1_{i\leq h\leq i+1}\cdot(1 - h + \lfloor h\rfloor)+\\
&\quad\quad\quad 1_{i-1\leq h\leq i}\cdot(h - \lfloor h\rfloor), \\
&\frac{\partial\varphi}{\partial h}\{h\}=\phi_{\lfloor h\rfloor+1}-\phi_{\lfloor h\rfloor}.
\end{aligned}
\end{equation}

The acquired zero-filled k-space $\bfB$ with undersampling mask $\bfM$ was normalized by 
$\frac{\|\bfM\|_1}{\|\bfB\|_\mathrm{F}}$.

To enable backpropagation to be carried out with limited GPU memory, we employ spatio-temporal equivariance of the convolution and exploit the fact that k-space is fully sampled in the readout dimension $k_x$ for Cartesian acquisitions.
Therefore, to draw a training sample, we perform random cropping of width $w_x$ and $w_t$ in dimensions $x$ and $t$ respectively and simulate Fourier encoding in $k_yk_z$ dimensions as illustrated in Fig.~\ref{fig:sketch_train}(a).
The network was implemented using the Tensorflow framework~\cite{abadi2016tensorflow}.
Fully-sampled and partial Fourier acquisition data from 11 healthy volunteers was used during training.

\subsubsection*{In Vivo Data Acquisition}
As illustrated in Fig.~\ref{fig:cs_acq_data}, we used 11 subject for network training, and 7 healthy subjects and 1 patient for evaluation. All in-vivo work was performed upon written informed consent of the subjects and according to local ethics regulations.

Training datasets comprised 4D flow data measured in the aorta of 11 healthy subjects, 9 of them fully sampled and 2 acquired with partial Fourier~\cite{cuppen1987reducing} (factor 0.75$\times$0.75).

For evaluation, data in the ascending aorta of 7 healthy subjects were acquired on a 3T Philips Ingenia system (Philips Healthcare, Best, the Netherlands) using a Cartesian 4-point referenced phase-contrast gradient-echo sequence with an encoding velocity $v_\mathrm{enc}=150\,cm/s$, a spatial resolution of $2.5$$\times$$2.5$$\times$$2.5\,mm^3$, $T_E = 3.3\,ms$, $T_R = 4.9\,ms$, $25$ cardiac phases and flip angle = 8$^\circ$. 
Exams for each of the 7 healthy subjects comprised a standard navigator-gated 2-fold accelerated parallel imaging~\cite{pruessmann1999sense} exam for reference, and a CS acquisition with an acceleration factor of $R$=$12.4~-~13.8$, using Cartesian pseudo-radial golden angle sampling pattern~\cite{winkelmann2006optimal} and data driven-respiratory motion detection, as in~\cite{walheim20195dflow}. Only data in expiration were kept for reconstruction as shown in Fig.~\ref{fig:cs_acq_data}. 

To evaluate reconstruction accuracy on pathological anatomy, 4D flow data was acquired in a single patient with dilation of the ascending aorta, combined aortic stenosis and regurgitation due to a bicuspid aortic valve on a 3T Philips Ingenia system (Philips Healthcare, Best, the Netherlands) using a navigator-gated 2-fold accelerated parallel imaging~\cite{pruessmann1999sense} scan.

A receiver coil with 28 channels was used for acquisition which were reduced to 5 channels using coil compression~\cite{zhang2013coil}. Coil sensitivity maps were estimated with ESPIRiT~\cite{uecker2014espirit}. Concomitant field correction was applied to the signal phase according to Bernstein et al.~\cite{bernstein1998concomitant} and eddy currents were corrected for with a third-order polynomial model fitted to stationary tissue \cite{busch2017image,walker1993semiautomated}. 

\subsubsection*{Evaluation}
We compared the proposed FlowVN to the state-of-the-art compressed sensing LLR-regularized~(\ref{eq:llr_reg}) reconstruction~\cite{zhang2015accelerating} and the variational network by Hammernik et al.~\cite{hammernik2018learning} which we refer to as HamVN.
The LLR implementation from the Berkeley advanced reconstruction toolbox (BART)~\cite{tamir2016generalized} was used with patch size $p$=$8$ and a maximum number of optimization iterations of 80.
The optimal value of regularization parameter $\lambda_\mathrm{LLR}$=$2.06$ was chosen via grid search to minimize the reconstructed flow field residual 
$\|\hat{\vectr{v}}-\vectr{v}^\ast\|_2$
averaged over the manually segmented aorta on the retrospectively 12$\times$ undersampled acquisition.
\vvv{Since the original VN~\cite{hammernik2018learning} was proposed for magnitude reconstruction of 2D and 3D data, we introduced the following modifications for the presented 4D flow evaluation: (i) 3D filters were grouped into 4 banks as in FlowVN (see Supplemental Algorithm~\ref{alg:varnet}), (ii) $\ell_1$-norm of reconstruction was optimized (i.e. equation~(\ref{eq:train_exp_loss}) with $\tau$$\rightarrow$+$\infty$ ).
	We refer to this architecture as ``HamVN'', the number of network layers, filters and control knots were the same as in FlowVN.
}

\subsubsection*{Retrospective Study}
For simulated retrospective undersampling experiments, we used 2$\times$ PI data and simulate a pseudo-radial Golden angle sampling pattern~\cite{winkelmann2006optimal} with acceleration factors of 6 to 22. 

For each undersampling factor we evaluated the nRMSE of the image magnitude, the relative error (RelErr) of velocity magnitudes inside the aorta and the angular error (AngErr) of the estimated velocity vectors:
\begin{equation}
\begin{aligned}
& \text{nRMSE}(\bfa, \bfa^\ast)=\sqrt{\sum^N_i \frac{(a_i-a^\star_i)^2}{N\text{max}_j (a^\star_j)^2}}, \\
& \text{RelErr}(\bfa, \bfa^\ast)=\frac{\|\bfa-\bfa^\ast\|_2}{\|\bfa^\ast\|_2}, \\
& \text{AngErr}(\mathbf{u}, \mathbf{v}) = \arccos\left( \frac{\left<\mathbf{u}, \mathbf{v}\right>}{\|\mathbf{u}\|_2\|\mathbf{v}\|_2} \right).
\end{aligned}
\end{equation}
\vvv{Additionally, we report the structural similarity index (SSIM)~\cite{wang2004image} with $\sigma_\text{SSIM}$=$1.5$ on the reconstructed magnitude images.}

\subsubsection*{Prospective Study}
Using manual aorta segmentations we compute flow over cross sections of the aorta by integrating velocity components projected onto the cross section normal.
The peak flow is then defined as the maximal flow over cardiac phases for a given cross section.
Moreover, we calculate peak through-plane velocity defined as maximum velocity projection across cross sections of the aorta over cardiac phases.

To quantify agreement with the reference 2$\times$ PI reconstruction, we performed Bland-Altman analysis~\cite{altman1983measurement} of peak flow and peak through-plane velocities. 

\subsection*{Data and code availability}
The code for the network training and inference used in this study and network weights are avaliable on CodeOcean together with volunteer data:
\url{https://codeocean.com/capsule/0115983/tree}~\cite{codeocn1}.
The code for analysis is available on CodeOcean:
\url{https://codeocean.com/capsule/2587940/tree}~\cite{codeocn2}.

\bibliographystyle{naturemag}
\bibliography{refsarx}

\begin{thebibliography}{10}
\expandafter\ifx\csname url\endcsname\relax
  \def\url#1{\texttt{#1}}\fi
\expandafter\ifx\csname urlprefix\endcsname\relax\def\urlprefix{URL }\fi
\providecommand{\bibinfo}[2]{#2}
\providecommand{\eprint}[2][]{\url{#2}}

\bibitem{markl20124d}
\bibinfo{author}{Markl, M.}, \bibinfo{author}{Frydrychowicz, A.},
  \bibinfo{author}{Kozerke, S.}, \bibinfo{author}{Hope, M.} \&
  \bibinfo{author}{Wieben, O.}
\newblock \bibinfo{title}{{4D} flow {MRI}}.
\newblock \emph{\bibinfo{journal}{J. Mag. Res. Imag.}}
  \textbf{\bibinfo{volume}{36}}, \bibinfo{pages}{1015--1036}
  (\bibinfo{year}{2012}).

\bibitem{feinberg1986halving}
\bibinfo{author}{Feinberg, D.}, \bibinfo{author}{Hale, J.},
  \bibinfo{author}{Watts, J.}, \bibinfo{author}{Kaufman, L.} \&
  \bibinfo{author}{Mark, A.}
\newblock \bibinfo{title}{Halving {MR} imaging time by conjugation:
  demonstration at 3.5 k{G}.}
\newblock \emph{\bibinfo{journal}{Radiol.}} \textbf{\bibinfo{volume}{161}},
  \bibinfo{pages}{527--531} (\bibinfo{year}{1986}).

\bibitem{szarf2006zero}
\bibinfo{author}{Szarf, G.} \emph{et~al.}
\newblock \bibinfo{title}{Zero filled partial fourier phase contrast {MR}
  imaging: in vitro and in vivo assessment}.
\newblock \emph{\bibinfo{journal}{J. Mag. Res. Imag.}}
  \textbf{\bibinfo{volume}{23}}, \bibinfo{pages}{42--49}
  (\bibinfo{year}{2006}).

\bibitem{walheim2019limitations}
\bibinfo{author}{Walheim, J.}, \bibinfo{author}{Gotschy, A.} \&
  \bibinfo{author}{Kozerke, S.}
\newblock \bibinfo{title}{On the limitations of partial fourier acquisition in
  phase-contrast {MRI} of turbulent kinetic energy}.
\newblock \emph{\bibinfo{journal}{Mag. Res. in Med.}}
  \textbf{\bibinfo{volume}{81}}, \bibinfo{pages}{514--523}
  (\bibinfo{year}{2019}).

\bibitem{pruessmann1999sense}
\bibinfo{author}{Pruessmann, K.}, \bibinfo{author}{Weiger, M.},
  \bibinfo{author}{Scheidegger, M.} \& \bibinfo{author}{Boesiger, P.}
\newblock \bibinfo{title}{{SENSE}: sensitivity encoding for fast {MRI}}.
\newblock \emph{\bibinfo{journal}{Mag. Res. in Med.}}
  \textbf{\bibinfo{volume}{42}}, \bibinfo{pages}{952--962}
  (\bibinfo{year}{1999}).

\bibitem{wiesinger2004electrodynamics}
\bibinfo{author}{Wiesinger, F.}, \bibinfo{author}{Boesiger, P.} \&
  \bibinfo{author}{Pruessmann, K.~P.}
\newblock \bibinfo{title}{Electrodynamics and ultimate {SNR} in parallel {MR}
  imaging}.
\newblock \emph{\bibinfo{journal}{Mag. Res. in Med.}}
  \textbf{\bibinfo{volume}{52}}, \bibinfo{pages}{376--390}
  (\bibinfo{year}{2004}).

\bibitem{lustig2007sparse}
\bibinfo{author}{Lustig, M.}, \bibinfo{author}{Donoho, D.} \&
  \bibinfo{author}{Pauly, J.~M.}
\newblock \bibinfo{title}{Sparse {MRI}: The application of compressed sensing
  for rapid {MR} imaging}.
\newblock \emph{\bibinfo{journal}{Mag. Res. in Med.}}
  \textbf{\bibinfo{volume}{58}}, \bibinfo{pages}{1182--1195}
  (\bibinfo{year}{2007}).

\bibitem{kim2012accelerated}
\bibinfo{author}{Kim, D.} \emph{et~al.}
\newblock \bibinfo{title}{Accelerated phase-contrast cine {MRI} using k-t
  {SPARSE-SENSE}}.
\newblock \emph{\bibinfo{journal}{Mag. Res. in Med.}}
  \textbf{\bibinfo{volume}{67}}, \bibinfo{pages}{1054--1064}
  (\bibinfo{year}{2012}).

\bibitem{valvano2017accelerating}
\bibinfo{author}{Valvano, G.} \emph{et~al.}
\newblock \bibinfo{title}{Accelerating {4D} flow {MRI} by exploiting low-rank
  matrix structure and hadamard sparsity}.
\newblock \emph{\bibinfo{journal}{Mag. Res. in Med.}}
  \textbf{\bibinfo{volume}{78}}, \bibinfo{pages}{1330--1341}
  (\bibinfo{year}{2017}).

\bibitem{bollache2018k}
\bibinfo{author}{Bollache, E.} \emph{et~al.}
\newblock \bibinfo{title}{k-t accelerated aortic 4{D} flow {MRI} in under two
  minutes: feasibility and impact of resolution, k-space sampling patterns, and
  respiratory navigator gating on hemodynamic measurements}.
\newblock \emph{\bibinfo{journal}{Mag. Res. in Med.}}
  \textbf{\bibinfo{volume}{79}}, \bibinfo{pages}{195--207}
  (\bibinfo{year}{2018}).

\bibitem{walheim20195dflow}
\bibinfo{author}{Walheim, J.}, \bibinfo{author}{Dillinger, H.} \&
  \bibinfo{author}{Kozerke, S.}
\newblock \bibinfo{title}{{Multipoint 5D Flow Cardiovascular Magnetic Resonance
  - Accelerated Cardiac- and Respiratory-Motion Resolved Mapping of Mean and
  Turbulent Velocities}}.
\newblock \emph{\bibinfo{journal}{J. Cardiovasc. Magn. Res.}}
  (\bibinfo{year}{2019}).

\bibitem{ma2019aortic}
\bibinfo{author}{Ma, L.~E.} \emph{et~al.}
\newblock \bibinfo{title}{Aortic {4D} flow {MRI} in 2 minutes using compressed
  sensing, respiratory controlled adaptive k-space reordering, and inline
  reconstruction}.
\newblock \emph{\bibinfo{journal}{Mag. Res. in Med.}}  (\bibinfo{year}{2019}).

\bibitem{rich2019bayesian}
\bibinfo{author}{Rich, A.} \emph{et~al.}
\newblock \bibinfo{title}{A bayesian approach for {4D} flow imaging of aortic
  valve in a single breath-hold}.
\newblock \emph{\bibinfo{journal}{Mag. Res. in Med.}}
  \textbf{\bibinfo{volume}{81}}, \bibinfo{pages}{811--824}
  (\bibinfo{year}{2019}).

\bibitem{zhang2015accelerating}
\bibinfo{author}{Zhang, T.}, \bibinfo{author}{Pauly, J.~M.} \&
  \bibinfo{author}{Levesque, I.~R.}
\newblock \bibinfo{title}{Accelerating parameter mapping with a locally low
  rank constraint}.
\newblock \emph{\bibinfo{journal}{Mag. Res. in Med.}}
  \textbf{\bibinfo{volume}{73}}, \bibinfo{pages}{655--661}
  (\bibinfo{year}{2015}).

\bibitem{hammernik2018learning}
\bibinfo{author}{Hammernik, K.} \emph{et~al.}
\newblock \bibinfo{title}{Learning a variational network for reconstruction of
  accelerated {MRI} data}.
\newblock \emph{\bibinfo{journal}{Mag. Res. in Med.}}
  \textbf{\bibinfo{volume}{79}}, \bibinfo{pages}{3055--3071}
  (\bibinfo{year}{2018}).

\bibitem{mardani2018deep}
\bibinfo{author}{Mardani, M.} \emph{et~al.}
\newblock \bibinfo{title}{Deep generative adversarial neural networks for
  compressive sensing {MRI}}.
\newblock \emph{\bibinfo{journal}{IEEE Trans. Med. Imag.}}
  \textbf{\bibinfo{volume}{38}}, \bibinfo{pages}{167--179}
  (\bibinfo{year}{2018}).

\bibitem{zhu2018image}
\bibinfo{author}{Zhu, B.}, \bibinfo{author}{Liu, J.~Z.},
  \bibinfo{author}{Cauley, S.~F.}, \bibinfo{author}{Rosen, B.~R.} \&
  \bibinfo{author}{Rosen, M.~S.}
\newblock \bibinfo{title}{Image reconstruction by domain-transform manifold
  learning}.
\newblock \emph{\bibinfo{journal}{Nature}} \textbf{\bibinfo{volume}{555}},
  \bibinfo{pages}{487} (\bibinfo{year}{2018}).

\bibitem{schlemper2017deep}
\bibinfo{author}{Schlemper, J.}, \bibinfo{author}{Caballero, J.},
  \bibinfo{author}{Hajnal, J.~V.}, \bibinfo{author}{Price, A.~N.} \&
  \bibinfo{author}{Rueckert, D.}
\newblock \bibinfo{title}{A deep cascade of convolutional neural networks for
  dynamic mr image reconstruction}.
\newblock \emph{\bibinfo{journal}{IEEE Trans. Med. Imag.}}
  \textbf{\bibinfo{volume}{37}}, \bibinfo{pages}{491--503}
  (\bibinfo{year}{2017}).

\bibitem{maier2019learning}
\bibinfo{author}{Maier, A.~K.} \emph{et~al.}
\newblock \bibinfo{title}{Learning with known operators reduces maximum error
  bounds}.
\newblock \emph{\bibinfo{journal}{Nat. Mach. Intel.}}
  \textbf{\bibinfo{volume}{1}}, \bibinfo{pages}{373--380}
  (\bibinfo{year}{2019}).

\bibitem{beck2009fast}
\bibinfo{author}{Beck, A.} \& \bibinfo{author}{Teboulle, M.}
\newblock \bibinfo{title}{A fast iterative shrinkage-thresholding algorithm for
  linear inverse problems}.
\newblock \emph{\bibinfo{journal}{SIAM J. Imag. Sc.}}
  \textbf{\bibinfo{volume}{2}}, \bibinfo{pages}{183--202}
  (\bibinfo{year}{2009}).

\bibitem{antun2019instabilities}
\bibinfo{author}{Antun, V.}, \bibinfo{author}{Renna, F.},
  \bibinfo{author}{Poon, C.}, \bibinfo{author}{Adcock, B.} \&
  \bibinfo{author}{Hansen, A.~C.}
\newblock \bibinfo{title}{On instabilities of deep learning in image
  reconstruction-does {AI} come at a cost?}
\newblock \emph{\bibinfo{journal}{arXiv preprint arXiv:1902.05300}}
  (\bibinfo{year}{2019}).

\bibitem{yang2017dagan}
\bibinfo{author}{Yang, G.} \emph{et~al.}
\newblock \bibinfo{title}{{DAGAN}: Deep de-aliasing generative adversarial
  networks for fast compressed sensing {MRI} reconstruction}.
\newblock \emph{\bibinfo{journal}{IEEE Trans. Med. Imag.}}
  \textbf{\bibinfo{volume}{37}}, \bibinfo{pages}{1310--1321}
  (\bibinfo{year}{2017}).

\bibitem{quan2018compressed}
\bibinfo{author}{Quan, T.}, \bibinfo{author}{Nguyen-Duc, T.} \&
  \bibinfo{author}{Jeong, W.}
\newblock \bibinfo{title}{Compressed sensing {MRI} reconstruction using a
  generative adversarial network with a cyclic loss}.
\newblock \emph{\bibinfo{journal}{IEEE Trans. Med. Imag.}}
  \textbf{\bibinfo{volume}{37}}, \bibinfo{pages}{1488--1497}
  (\bibinfo{year}{2018}).

\bibitem{narnhofer2019inverse}
\bibinfo{author}{Narnhofer, D.}, \bibinfo{author}{Hammernik, K.},
  \bibinfo{author}{Knoll, F.} \& \bibinfo{author}{Pock, T.}
\newblock \bibinfo{title}{Inverse {GANs} for accelerated {MRI} reconstruction}.
\newblock In \emph{\bibinfo{booktitle}{Wav. and Spars.}}, vol.
  \bibinfo{volume}{11138}, \bibinfo{pages}{111381A} (\bibinfo{year}{2019}).

\bibitem{zhang2010magnetic}
\bibinfo{author}{Zhang, S.}, \bibinfo{author}{Block, K.} \&
  \bibinfo{author}{Frahm, J.}
\newblock \bibinfo{title}{Magnetic resonance imaging in real time: advances
  using radial {FLASH}}.
\newblock \emph{\bibinfo{journal}{J. Mag. Res. Imag.}}
  \textbf{\bibinfo{volume}{31}}, \bibinfo{pages}{101--109}
  (\bibinfo{year}{2010}).

\bibitem{landweber1951iteration}
\bibinfo{author}{Landweber, L.}
\newblock \bibinfo{title}{An iteration formula for {F}redholm integral
  equations of the first kind}.
\newblock \emph{\bibinfo{journal}{Am. J. of Math.}}
  \textbf{\bibinfo{volume}{73}}, \bibinfo{pages}{615--624}
  (\bibinfo{year}{1951}).

\bibitem{liu2016learning}
\bibinfo{author}{Liu, Y.} \& \bibinfo{author}{Lew, M.~S.}
\newblock \bibinfo{title}{Learning relaxed deep supervision for better edge
  detection}.
\newblock In \emph{\bibinfo{booktitle}{CVPR}}, \bibinfo{pages}{231--240}
  (\bibinfo{year}{2016}).

\bibitem{ravishankar2010mr}
\bibinfo{author}{Ravishankar, S.} \& \bibinfo{author}{Bresler, Y.}
\newblock \bibinfo{title}{{MR} image reconstruction from highly undersampled
  k-space data by dictionary learning}.
\newblock \emph{\bibinfo{journal}{IEEE Trans. Med. Imag.}}
  \textbf{\bibinfo{volume}{30}}, \bibinfo{pages}{1028--1041}
  (\bibinfo{year}{2010}).

\bibitem{caballero2014dictionary}
\bibinfo{author}{Caballero, J.}, \bibinfo{author}{Price, A.~N.},
  \bibinfo{author}{Rueckert, D.} \& \bibinfo{author}{Hajnal, J.~V.}
\newblock \bibinfo{title}{Dictionary learning and time sparsity for dynamic
  {MR} data reconstruction}.
\newblock \emph{\bibinfo{journal}{IEEE Trans. Med. Imag.}}
  \textbf{\bibinfo{volume}{33}}, \bibinfo{pages}{979--994}
  (\bibinfo{year}{2014}).

\bibitem{lee2017deep}
\bibinfo{author}{Lee, D.}, \bibinfo{author}{Yoo, J.} \& \bibinfo{author}{Ye,
  J.~C.}
\newblock \bibinfo{title}{Deep artifact learning for compressed sensing and
  parallel {MRI}}.
\newblock \emph{\bibinfo{journal}{arXiv preprint arXiv:1703.01120}}
  (\bibinfo{year}{2017}).

\bibitem{lecun1988theoretical}
\bibinfo{author}{LeCun, Y.}, \bibinfo{author}{Touresky, D.},
  \bibinfo{author}{Hinton, G.} \& \bibinfo{author}{Sejnowski, T.}
\newblock \bibinfo{title}{A theoretical framework for back-propagation}.
\newblock In \emph{\bibinfo{booktitle}{Proc. Connect. Mod.}},
  vol.~\bibinfo{volume}{1}, \bibinfo{pages}{21--28} (\bibinfo{year}{1988}).

\bibitem{abadi2016tensorflow}
\bibinfo{author}{Abadi, M.} \emph{et~al.}
\newblock \bibinfo{title}{Tensorflow: A system for large-scale machine
  learning}.
\newblock In \emph{\bibinfo{booktitle}{Proc. {USENIX}}},
  \bibinfo{pages}{265--283} (\bibinfo{year}{2016}).

\bibitem{domke2012generic}
\bibinfo{author}{Domke, J.}
\newblock \bibinfo{title}{Generic methods for optimization-based modeling}.
\newblock In \emph{\bibinfo{booktitle}{Art. Int. and Stat.}},
  \bibinfo{pages}{318--326} (\bibinfo{year}{2012}).

\bibitem{sun2016deep}
\bibinfo{author}{Sun, J.}, \bibinfo{author}{Li, H.}, \bibinfo{author}{Xu, Z.}
  \emph{et~al.}
\newblock \bibinfo{title}{Deep {ADMM}-{N}et for compressive sensing {MRI}}.
\newblock In \emph{\bibinfo{booktitle}{NIPS}}, \bibinfo{pages}{10--18}
  (\bibinfo{year}{2016}).

\bibitem{jin2017deep}
\bibinfo{author}{Jin, K.~H.}, \bibinfo{author}{McCann, M.~T.},
  \bibinfo{author}{Froustey, E.} \& \bibinfo{author}{Unser, M.}
\newblock \bibinfo{title}{Deep convolutional neural network for inverse
  problems in imaging}.
\newblock \emph{\bibinfo{journal}{IEEE Trans. on Imag. Proc.}}
  \textbf{\bibinfo{volume}{26}}, \bibinfo{pages}{4509--4522}
  (\bibinfo{year}{2017}).

\bibitem{vishnevskiy2018image}
\bibinfo{author}{Vishnevskiy, V.}, \bibinfo{author}{Sanabria, S.~J.} \&
  \bibinfo{author}{Goksel, O.}
\newblock \bibinfo{title}{Image reconstruction via variational network for
  real-time hand-held sound-speed imaging}.
\newblock In \emph{\bibinfo{booktitle}{Intl. W. on Mach. Learn. for Med. Imag.
  Recon.}}, \bibinfo{pages}{120--128} (\bibinfo{year}{2018}).

\bibitem{vishnevskiy2019deep}
\bibinfo{author}{Vishnevskiy, V.}, \bibinfo{author}{Rau, R.} \&
  \bibinfo{author}{Goksel, O.}
\newblock \bibinfo{title}{Deep variational networks with exponential weighting
  for learning computed tomography}.
\newblock In \emph{\bibinfo{booktitle}{MICCAI}}, \bibinfo{pages}{310--318}
  (\bibinfo{year}{2019}).

\bibitem{cuppen1987reducing}
\bibinfo{author}{Cuppen, J.} \& \bibinfo{author}{van Est, A.}
\newblock \bibinfo{title}{Reducing {MR} imaging time by one-sided
  reconstruction}.
\newblock \emph{\bibinfo{journal}{Mag. Res. Imag.}}
  \textbf{\bibinfo{volume}{5}}, \bibinfo{pages}{526--527}
  (\bibinfo{year}{1987}).

\bibitem{winkelmann2006optimal}
\bibinfo{author}{Winkelmann, S.}, \bibinfo{author}{Schaeffter, T.},
  \bibinfo{author}{Koehler, T.}, \bibinfo{author}{Eggers, H.} \&
  \bibinfo{author}{Doessel, O.}
\newblock \bibinfo{title}{An optimal radial profile order based on the {G}olden
  {R}atio for time-resolved {MRI}}.
\newblock \emph{\bibinfo{journal}{IEEE Trans. Med. Imag.}}
  \textbf{\bibinfo{volume}{26}}, \bibinfo{pages}{68--76}
  (\bibinfo{year}{2006}).

\bibitem{zhang2013coil}
\bibinfo{author}{Zhang, T.}, \bibinfo{author}{Pauly, J.~M.},
  \bibinfo{author}{Vasanawala, S.~S.} \& \bibinfo{author}{Lustig, M.}
\newblock \bibinfo{title}{Coil compression for accelerated imaging with
  cartesian sampling}.
\newblock \emph{\bibinfo{journal}{Mag. Res. in Med.}}
  \textbf{\bibinfo{volume}{69}}, \bibinfo{pages}{571--582}
  (\bibinfo{year}{2013}).

\bibitem{uecker2014espirit}
\bibinfo{author}{Uecker, M.} \emph{et~al.}
\newblock \bibinfo{title}{{ESPIRiT}~--- an eigenvalue approach to
  autocalibrating parallel {MRI}: where {SENSE} meets {GRAPPA}}.
\newblock \emph{\bibinfo{journal}{Mag. Res. in Med.}}
  \textbf{\bibinfo{volume}{71}}, \bibinfo{pages}{990--1001}
  (\bibinfo{year}{2014}).

\bibitem{bernstein1998concomitant}
\bibinfo{author}{Bernstein, M.~A.} \emph{et~al.}
\newblock \bibinfo{title}{Concomitant gradient terms in phase contrast {MR}:
  analysis and correction}.
\newblock \emph{\bibinfo{journal}{Mag. Res. in Med.}}
  \textbf{\bibinfo{volume}{39}}, \bibinfo{pages}{300--308}
  (\bibinfo{year}{1998}).

\bibitem{busch2017image}
\bibinfo{author}{Busch, J.}, \bibinfo{author}{Giese, D.} \&
  \bibinfo{author}{Kozerke, S.}
\newblock \bibinfo{title}{Image-based background phase error correction in 4{D}
  flow {MRI} revisited}.
\newblock \emph{\bibinfo{journal}{J. of MRI}} \textbf{\bibinfo{volume}{46}},
  \bibinfo{pages}{1516--1525} (\bibinfo{year}{2017}).

\bibitem{walker1993semiautomated}
\bibinfo{author}{Walker, P.~G.} \emph{et~al.}
\newblock \bibinfo{title}{Semiautomated method for noise reduction and
  background phase error correction in {MR} phase velocity data}.
\newblock \emph{\bibinfo{journal}{J. Mag. Res. Imag.}}
  \textbf{\bibinfo{volume}{3}}, \bibinfo{pages}{521--530}
  (\bibinfo{year}{1993}).

\bibitem{tamir2016generalized}
\bibinfo{author}{Tamir, J.~I.}, \bibinfo{author}{Ong, F.},
  \bibinfo{author}{Cheng, J.~Y.}, \bibinfo{author}{Uecker, M.} \&
  \bibinfo{author}{Lustig, M.}
\newblock \bibinfo{title}{Generalized magnetic resonance image reconstruction
  using the {B}erkeley advanced reconstruction toolbox}.
\newblock In \emph{\bibinfo{booktitle}{ISMRM W. on Dat. Sampl. and Imag.
  Recon.}} (\bibinfo{year}{2016}).

\bibitem{wang2004image}
\bibinfo{author}{Wang, Z.}, \bibinfo{author}{Bovik, A.~C.},
  \bibinfo{author}{Sheikh, H.~R.} \& \bibinfo{author}{Simoncelli, E.~P.}
\newblock \bibinfo{title}{Image quality assessment: from error visibility to
  structural similarity}.
\newblock \emph{\bibinfo{journal}{IEEE Trans. on Imag. Proc.}}
  \textbf{\bibinfo{volume}{13}}, \bibinfo{pages}{600--612}
  (\bibinfo{year}{2004}).

\bibitem{altman1983measurement}
\bibinfo{author}{Altman, D.~G.} \& \bibinfo{author}{Bland, J.~M.}
\newblock \bibinfo{title}{Measurement in medicine: the analysis of method
  comparison studies}.
\newblock \emph{\bibinfo{journal}{J. R. Stat. Soc.}}
  \textbf{\bibinfo{volume}{32}}, \bibinfo{pages}{307--317}
  (\bibinfo{year}{1983}).

\bibitem{codeocn1}
\bibinfo{author}{Vishnevskiy, V.}, \bibinfo{author}{Walheim, J.} \&
  \bibinfo{author}{Kozerke, S.}
\newblock \bibinfo{title}{{FlowVN}: deep variational network for rapid 4{D}
  flow {MRI} reconstruction}.
\newblock \emph{\bibinfo{journal}{CodeOcean}}
  \urlprefix\url{https://codeocean.com/capsule/0115983/tree}.
\newblock \bibinfo{note}{(2020)}.

\bibitem{codeocn2}
\bibinfo{author}{Vishnevskiy, V.}, \bibinfo{author}{Walheim, J.} \&
  \bibinfo{author}{Kozerke, S.}
\newblock \bibinfo{title}{{FlowVN}: analysis}.
\newblock \emph{\bibinfo{journal}{CodeOcean}} \urlprefix\url{CodeOcean
  \url{https://codeocean.com/capsule/5994453/tree}}.
\newblock \bibinfo{note}{(2020)}.

\end{thebibliography}
\subsection*{Acknowledgements}
The authors acknowledge funding from the European Unions Horizon 2020 research and innovation program under grant agreement No 668039 and under EuroStars UNIFORM as well as funding of the Platform for Advanced Scientific Computing of the Council of the Federal Institutes of Technology (ETH Board), Switzerland.

\subsection*{Author contributions}
J.W., V.V. and S.K. conceived the study.
V.V. implemented machine learning reconstruction algorithms. 
J.W. conducted MR acquisition experiments and data preprocessing. 
J.W. and V.V. analyzed experimental data under supervision of S.K. 
All authors discussed the results and contributed to writing the manuscript.

\subsection*{Competing interests}
The authors declare no competing interests.

\newpage
\section{Supplementary Material}
\begin{algorithm*}[h!]
	\caption{Proposed variational reconstruction network model FlowVN. }
	\label{alg:varnet}
	\begin{tabular}{l}
		{Input:  $\bfB\in\bbCo^{\Ns\times\Nt}$~---~zero-filled k-space samples, $\bfM\in\{0,1\}^{\Ns\times\Nt}$~---~undersampling mask}\\
		{Parameters: $\Theta = \{
			\boldsymbol{\phi}^{(k)}_\text{ud},
			\boldsymbol{\phi}^{(k)}_\text{ur}, 
			\boldsymbol{\phi}^{(k)}_\text{d},\boldsymbol{\phi}^{(k)}_{\text{r},in},  \matr{D}_{in}^{(k)}, \alpha^{(k)}, \alpha^{(0)}\}_{i=1,\dots,N_f,\; k=1,\dots,K},\; n=1,\dots,4$}\\
		{$\overline{\bfM}\leftarrow\frac1{\Ns\Nt}\sum_{i\leq\Ns j\leq\Nt}M_{ij}$}\\
		{$\bfRho^{(0)} \leftarrow \alpha^{(0)}\matr{E}\conj\bfB,\quad\vectr{S}^{(0)}\leftarrow\vectr{0}$}\\
		{\texttt{\textbf{for}} $k:=0$ to $K-1$}\\
		{\quad $\vectr{G}^{(k)}\leftarrow 
			\overbrace{\varphi_\mathrm{ud}^{(k)}\{\overline{\bfM}\} \matr{E}\conj\left( \bfM\odot  \varphi_\text{d}^{(k)}\left\{\bfM\odot \left(\matr{E}\bfRho^{(k)} - \bfB\right) \right\}\right)}^\text{data consistency term} + 
			\overbrace{	\varphi_\mathrm{ur}^{(k)}\{\overline{\bfM}\}		\sum\limits_{i\leq N_\mathrm{f}, n\leq4}\left(\matr{D}_{in}^{(k)}\right)\tran\,\varphi^{(k)}_{\text{r}, in} \left\{  \matr{D}_{in}^{(k)}\mathrm{vec}\left(\bfRho^{(k)}\right)\right\}}^\text{regularization term}$}\\
		{\quad $\vectr{S}^{(k+1)}\leftarrow \alpha^{(k+1)}\vectr{S}^{(k)} + \vectr{G}^{(k)}$}\\
		{\quad $\bfRho^{(k+1)}\leftarrow \bfRho^{(k)} - \vectr{S}^{(k+1)}$}\\
		{Output: reconstructed image $\mathcal{V}(\bfB, \bfM; \,\Theta) \defeq \bfRho^{(K)}$}\\
	\end{tabular}
\end{algorithm*}

\begin{table*}[h!]
	\footnotesize
	\caption{
		Comparison of reconstruction errors ($\pm$ standard deviation) of 7 retrospectively undersampled acquisitions of healthy subjects for different acceleration factors $R$.
		For each row the best performing method is highlighted in bold. 
		\vvv{P-values are obtained by comparing error metrics distributions of LLR and corresponding VN architectures using the two-sided Wilcoxon signed-rank test.
		}
	}
	\label{tbl:retro_errors}
	\begin{threeparttable}
		\tiny
		\begin{minipage}[t]{0.05\linewidth}
			\begin{tabular}{l}
				\headrow
				~\\
				\textbf{~} \\
				$R$=6 \\
				{\scriptsize \,\,\vvv{p-val.}}\\
				$R$=8 \\
				{\scriptsize \,\,\vvv{p-val.}}\\
				$R$=10\\
				{\scriptsize \,\,\vvv{p-val.}}\\
				$R$=12\\
				{\scriptsize \,\,\vvv{p-val.}}\\
				$R$=14 \\
				{\scriptsize \,\,\vvv{p-val.}}\\
				$R$=16 \\
				{\scriptsize \,\,\vvv{p-val.}}\\
				$R$=18 \\
				{\scriptsize \,\,\vvv{p-val.}}\\
				$R$=20 \\
				{\scriptsize \,\,\vvv{p-val.}}\\
				$R$=22 \\
				{\scriptsize \,\,\vvv{p-val.}}\\
				\hline
			\end{tabular}
		\end{minipage}
		\begin{minipage}[t]{0.25\linewidth}
			\begin{tabular}{|ccc|}
				\headrow
				\multicolumn{3}{c}{
					{Image Magnitude nRMSE [\%]}
				}\\
				\textbf{LLR} & \textbf{HamVN} & ~\textbf{FlowVN}\\
				1.9$\pm$0.3   & 2.3$\pm$0.2   & \textbf{1.7}$\pm$0.2   \\
				~& \vvv{\scriptsize $<$0.0001} & \vvv{\scriptsize 0.0017} \\
				2.0$\pm$0.3   & 2.5$\pm$0.2   & \textbf{1.9}$\pm$0.2   \\
				~& \vvv{\scriptsize $<$0.0001} & \vvv{\scriptsize 0.0076} \\
				\textbf{2.2}$\pm$0.3   & 2.8$\pm$0.3   & \textbf{2.2}$\pm$0.2   \\
				~& \vvv{\scriptsize $<$0.0001} & \vvv{\scriptsize 0.1418} \\
				2.5$\pm$0.3   & 3.0$\pm$0.3   & \textbf{2.4}$\pm$0.2   \\
				~& \vvv{\scriptsize $<$0.0001} & \vvv{\scriptsize 0.0256} \\
				2.8$\pm$0.3   & 3.3$\pm$0.3   & \textbf{2.6}$\pm$0.3   \\
				~& \vvv{\scriptsize $<$0.0001} & \vvv{\scriptsize $<$0.0001} \\
				3.3$\pm$0.4   & 3.6$\pm$0.3   & \textbf{2.9}$\pm$0.3   \\
				~& \vvv{\scriptsize $<$0.0001} & \vvv{\scriptsize $<$0.0001} \\
				3.6$\pm$0.4   & 3.8$\pm$0.4   & \textbf{3.0}$\pm$0.3   \\
				~& \vvv{\scriptsize $<$0.0001} & \vvv{\scriptsize $<$0.0001} \\
				3.8$\pm$0.4   & 4.0$\pm$0.4   & \textbf{3.2}$\pm$0.3   \\
				~& \vvv{\scriptsize 0.0211} & \vvv{\scriptsize $<$0.0001} \\
				4.2$\pm$0.5   & 4.1$\pm$0.4   & \textbf{3.4}$\pm$0.3   \\
				~& \vvv{\scriptsize 0.9551} & \vvv{\scriptsize $<$0.0001} \\
				\hline
			\end{tabular}
		\end{minipage}
		\begin{minipage}[t]{0.25\linewidth}
			\begin{tabular}{ccc|}
				\headrow
				\multicolumn{3}{c}{
					{Velocity Magnitude RelErr [\%]}
				}\\
				\textbf{LLR} & \textbf{HamVN} & ~\textbf{FlowVN}\\
				9.4$\pm$1.9	 & 12.8$\pm$2.5	 & \textbf{8.7}$\pm$2.4	\\
				~& \vvv{\scriptsize $<$0.0001} & \vvv{\scriptsize 0.0225} \\
				10.5$\pm$2.3	 & 13.5$\pm$2.8	 & \textbf{10.0}$\pm$2.8	\\
				~& \vvv{\scriptsize $<$0.0001} & \vvv{\scriptsize 0.0790} \\				
				11.8$\pm$2.2	 & 14.9$\pm$3.0	 & \textbf{11.4}$\pm$2.8	\\
				~& \vvv{\scriptsize $<$0.0001} & \vvv{\scriptsize 0.0385} \\				
				13.2$\pm$2.4	 & 16.4$\pm$2.9	 & \textbf{12.4}$\pm$2.4	\\
				~& \vvv{\scriptsize $<$0.0001} & \vvv{\scriptsize 0.0003} \\
				15.1$\pm$2.5	 & 18.0$\pm$3.5	 & \textbf{14.0}$\pm$2.9	\\
				~& \vvv{\scriptsize $<$0.0001} & \vvv{\scriptsize $<$0.0001} \\				
				17.4$\pm$2.3	 & 20.2$\pm$3.2	 & \textbf{15.5}$\pm$3.3	\\
				~& \vvv{\scriptsize $<$0.0001} & \vvv{\scriptsize $<$0.0001} \\				
				19.0$\pm$2.7	 & 22.4$\pm$2.9	 & \textbf{16.9}$\pm$2.9	\\
				~& \vvv{\scriptsize $<$0.0001} & \vvv{\scriptsize $<$0.0001} \\				
				19.9$\pm$3.4	 & 23.1$\pm$4.1	 & \textbf{17.5}$\pm$3.4	\\
				~& \vvv{\scriptsize $<$0.0001} & \vvv{\scriptsize $<$0.0001} \\				
				21.9$\pm$2.8	 & 25.3$\pm$3.6	 & \textbf{19.2}$\pm$3.3	\\
				~& \vvv{\scriptsize $<$0.0001} & \vvv{\scriptsize $<$0.0001} \\				
				\hline
			\end{tabular}
		\end{minipage}
		\begin{minipage}[t]{0.25\linewidth}
			\begin{tabular}{ccc}
				\headrow
				\multicolumn{3}{c}{
					{Mean Flow AngErr [deg]}
				}\\
				\textbf{LLR} & \textbf{HamVN} & ~\textbf{FlowVN}\\
				7.0$\pm$1.8	 & 9.6$\pm$2.6	 & \textbf{6.9}$\pm$1.9	\\
				~& \vvv{\scriptsize $<$0.0001} & \vvv{\scriptsize 0.1956} \\				
				\textbf{7.9}$\pm$2.0	 & 10.5$\pm$2.9	 & 8.0$\pm$2.3	\\
				~& \vvv{\scriptsize $<$0.0001} & \vvv{\scriptsize 0.2228} \\				
				\textbf{8.9}$\pm$2.2	 & 11.5$\pm$3.1	 & 9.2$\pm$2.5	\\
				~& \vvv{\scriptsize $<$0.0001} & \vvv{\scriptsize 0.0204} \\				
				\textbf{10.2}$\pm$2.7	 & 12.7$\pm$3.4	 & 10.3$\pm$2.8	\\
				~& \vvv{\scriptsize $<$0.0001} & \vvv{\scriptsize 0.0225} \\				
				\textbf{11.3}$\pm$2.9	 & 13.6$\pm$3.8	 & 11.4$\pm$3.2	\\
				~& \vvv{\scriptsize $<$0.0001} & \vvv{\scriptsize 0.1166} \\				
				\textbf{12.2}$\pm$3.1	 & 14.3$\pm$4.0	 & 12.3$\pm$3.4	\\
				~& \vvv{\scriptsize $<$0.0001} & \vvv{\scriptsize 0.7785} \\				
				13.3$\pm$3.4	 & 15.3$\pm$4.2	 & \textbf{13.0}$\pm$3.6	\\
				~& \vvv{\scriptsize $<$0.0001} & \vvv{\scriptsize 0.4799} \\				
				13.7$\pm$3.7	 & 15.5$\pm$4.4	 & \textbf{13.6}$\pm$3.9	\\
				~& \vvv{\scriptsize $<$0.0001} & \vvv{\scriptsize 0.0073} \\				
				14.8$\pm$3.6	 & 16.1$\pm$4.5	 & \textbf{14.3}$\pm$3.8	\\
				~& \vvv{\scriptsize $<$0.0001} & \vvv{\scriptsize 0.0005} \\				
				\hline
			\end{tabular}
		\end{minipage}
		\hfill
	\end{threeparttable}
\end{table*}

\begin{table*}[h!]
	\caption{
		\vvv{
			Quantitative prospective comparison of VN architectures and modifications proposed in this paper using data from 7 healthy volunteers ($12.4\leq R\leq13.8$).
			Peak velocities and peak flow are assessed over manually segmented aorta slices and reported relative to average values of the corresponding reference scan ($\pm$ standard deviation).
			P-values are obtained by comparing error metrics distributions of reference PI data and the corresponding reconstruction method using the two-sided Wilcoxon signed-rank test.
			HamVN-3D performs  3D filtering of $yzt$ dimensional data (assuming that data is fully sampled in the $x$ dimension).
			``HamVN-3D, single coil'' was trained to reconstruct each coil image separately and then to be combined using provided sensitivity maps.
			Modifications are accumulated as presented and, eventually, yield the FlowVN configuration.}
	}
	\label{tbl:pro_ablation}
	\tiny
	\begin{tabular}{l|cr|cr}
		\textbf{Method} & \textbf{Peak Velocity Error}[\%] & p-value  & \textbf{Peak Flow Error}[\%] & p-value \\ \hline
		HamVN-3D, single coil		& 19.38$\pm$19.01 & $<$0.0001 & 21.37$\pm$15.11& $<$0.0001 \\ 
		HamVN-3D					&  6.44$\pm$13.64 & $<$0.0001 & 3.73 $\pm$ 12.12& 0.0020 \\
		HamVN		             & 3.64$\pm$10.10     & $<$0.0001 & 1.90$\pm$9.69   & 0.3345\\
		+linear activation functions & 3.18$\pm$10.35 & 0.0002 & 1.65$\pm$10.70 & 0.3833 \\
		+unroll GD with momentum   & 2.95$\pm$9.89    & 0.0004 & 1.47$\pm$9.82  & 0.4749\\
		+data activation functions 	 & 2.29$\pm$10.12 & 0.0213 & 0.63$\pm$9.74 & 0.7957\\    
		+US-depending modulation & 1.98$\pm$8.43      & 0.0645 &  0.47$\pm$10.30 & 0.9272\\    
		+exp. weighting (FlowVN)& \textbf{1.59}$\pm$9.65 & 0.0875 &  \textbf{0.05}$\pm$9.79 & 0.9550\\    
	\end{tabular}
	
\end{table*}

\end{document}